\documentclass[reprint,amsmath,amssymb,aps,prl,superscriptaddress]{revtex4-2}
\usepackage[latin9]{inputenc}
\usepackage{verbatim}
\usepackage{amsmath}
\usepackage{amssymb}

\makeatletter
\usepackage{graphicx}
\usepackage{dcolumn}
\usepackage{bm}

\usepackage{xcolor}

\usepackage{graphicx}
\usepackage{ragged2e} 
\newcommand{\di}{i}          
\renewcommand{\vec}[1]{\mathbf{#1}}

\makeatother

\begin{document}
\title{Absence of Friedel oscillations in the entanglement entropy profile of one-dimensional intrinsically gapless topological phases}
\author{Shun-Chiao Chang}
\author{Pavan Hosur}
\affiliation{Department of Physics, University of Houston, Houston, Texas 77204, USA}
\affiliation{Texas Center for Superconductivity, Houston, Texas 77204, USA
}

\begin{abstract}
Topological quantum matter is typically associated with gapped phases and edge modes protected by the bulk gap. In contrast, recent work (Phys. Rev. B 104, 075132) proposed intrinsically gapless topological phases that, in one dimension, carry protected edge modes only when the bulk is a gapless Luttinger liquid. The edge modes of such a topological Luttinger liquid (TLL) descend from a nonlocal string order that is forbidden in gapped phases and whose precise form depends on the symmetry class of the system.
In this work, we propose a powerful and unbiased entanglement-based smoking gun signature of the TLL. In particular, we show that the entanglement entropy profile of a TLL lacks Friedel oscillations that are invariably present in other gapless one dimensional phases such as ordinary Luttinger liquids, and argue that their absence is closely related to a long-ranged string order which is an intrinsic property of the TLL. Crucially, such a diagnostic is more robust against numerical errors and relatively easier to measure in experiments as it relies on the entanglement entropy rather than entanglement spectrum, unlike the entanglement-based diagnostics of gapped topological phases in one dimension.

\end{abstract}
\pacs{Valid PACS appear here}

\maketitle


	\par $Introduction.$-- Topological quantum matter is defined by the obstruction to smooth deformation  into a product state, and is usually associated with gapped phases accompanied by gapless edge modes. Symmetry protected topological phases (SPTs) are a subset of gapped topological phases that are only non-trivial in the presence of certain global symmetries \cite{enrichref1,enrichref2,enrichref3,enrichref4,enrichref5,enrichref6,enrichref8,mainref1,mainref2}. 
In one dimension (1D), the symmetry and the bulk gap together protect localized zero modes on the edge even if the presence of interactions. Interestingly, the last decade has revealed many instances where the edge modes survive even when the bulk gap closes \cite{mainref4,mainref5,mainref6,mainref7,mainref8,mainref9,mainref10,mainref11,mainref12,mainref13,mainref14,mainref15,mainref16,mainref17,mainref18,mainref19,mainref20,mainref21,mainref22,mainref23,mainref24,mainref26,mainref27,mainref28,mainref29,mainref30,mainref31,mainref32,mainref33,mainref34}.

\par On the other hand, a recent breakthrough demonstrated the existence of intrinsically gapless SPTs \cite{main}. Such a phase is defined by the appearance of a quantum anomaly in its bulk low energy, local theory and can emerge when the microscopic symmetry is on-site. In 1D, appears as a 
Luttinger liquid. However, it differs from a trivial or ordinary Luttinger liquid (OLL) -- and forms a topological Luttinger liquid (TLL) instead -- through the presence of a long ranged string order and protected edge modes. These features are reminiscent of gapped SPTs, but they cannot be viewed as remnants of a gapped SPT whose bulk gap has closed \cite{DHLeespt}, resulting in what we refer to as gapless SPTs. On the other hand, unlike gapped SPTs where the string order and edge modes rely on a bulk gap, these topological features in a TLL \emph{require} the bulk to be gapless, and disappear if the TLL transitions into a gapped insulator. In this work, we reserve the term TLL for intrinsically gapless SPTs only.

\par For the topological phases, entanglement properties such as the entanglement spectrum and the entanglement entropy \cite{exhboseref50,exhboseref51,EEmore1,EEmore2,EEmore3,EEmore4,EEmore5,EEmore6,EEmore7,EEmore9,EEmore10,EEmore11,EEmore12,EEmore13,disconnectedentropy} provide powerful alternate diagnostics of the topological phase. The former reveals the topology through gapless modes and degeneracies and the latter, through universal sub-leading corrections to the area law. These diagnostics are unbiased as they do not require knowledge of the precise form of the string order parameter. This inspires the question, ``what entanglement properties, if any, are unique to TLLs and distinguish them from other gapped and gapless phases in 1D?"
 
 \par In this work, we show that TLLs have a distinct entanglement entropy (EE) profile $S_{vN}(j)$ as a function of the subsystem size $j$. While it was shown to grow logarithmically with $j$ as expected for a gapless phase \cite{main}, we find that it conspicuously lacks the Friedel oscillations that invariably occur in the EE profiles of OLLs \cite{eeprofilekf} and gapless SPTs \cite{DHLeespt}. 
Thus, we propose the absence of oscillations in the EE profile as a diagnostic for distinguishing TLLs from other gapless phases in 1D.
 
All computations were performed using the density matrix renormalization group (DMRG) \cite{itensor}. We mainly study two models: (i) an Ising-Hubbard chain of length $L=64$ with $222$ sweeps and final maximum bond dimension $\chi\approx500$ resulting in truncation error $\epsilon\approx10^{-8}$; (ii) a spin-1 doped Haldane chain, implemented as a t-J ladder with $L=40$ rungs, with $\chi\approx4000$ resulting in $\epsilon\approx10^{-9}$. We also present brief results on two other realizations of TLLs $L=50$ described later.


\par \emph{Ising-Hubbard}.-- The Ising-Hubbard model we use is given by  $H_1=H_{\textrm{Ising}}+H_{\textrm{Hub}}+H_{\textrm{inv}}$ where
\begin{eqnarray}
	H_{\textrm{Ising}}&=&\sum_jJ_zS_j^zS_{j+1}^z+h_xS_j^x \label{eq:Ising} \\
	H_{\textrm{Hub}}&=&-t\sum_{j,s}A_j^s+U\sum_jn_{j\uparrow}n_{j\downarrow}-\mu N \label{eq:hub} \\
	H_{\textrm{inv}}&=&-t_c\sum_{j,s}(\di c^\dag_{j+2,s}c_{j,s}+h.c.) \label{eq:inversion} 
\end{eqnarray}
with $c_{j,s}^\dagger$ creating a fermion at site $j$ with spin $s$, $n_{js}=c^\dag_{j,s}c_{j,s}$, $A^s=c^\dag_{j+1,s}c_{j,s}+h.c.$ and $S^\alpha_j= \frac{1}{2} c^\dag_{j,s} \sigma^\alpha_{s,{s}'} c_{j,{s}'}$. At $t_c=0$, $H_1$ reduces to the pure Ising-Hubbard model studied in Ref.~\cite{main}.
The key symmetry of this model is  $\mathbb{Z}_4=\{1,R_x,R_x^2,R_x^3\}$, where $R_x=\exp\left(i\pi \sum_jS_j^x\right)$ flips $S_z$ and squares to fermion parity: $R_x^2=(-1)^N\equiv P$. At $t_c=0$ and $J_z>0$, the ground state at half-filling is a Mott insulator -- either an Ising anti-ferromagnet at small $h_x$ or a paramagnet at large $h_x$. Tuning $U$ and $\mu$ drives the paramagnetic insulator into an $S^x$-polarized  OLL, whereas the Ising Mott insulator transitions into a TLL protected by the $\mathbb{Z}_4$ symmetry. In this work, we add second nearest neighbor imaginary hopping $H_{\textrm{inv}}$ that breaks inversion symmetry but preserves $R_x$. Since the TLL is protected by $R_x$, it remains robust up to moderate strengths of this hopping.

The TLL can be viewed as a hidden symmetry breaking state in which arbitrary number of holes or doublons are inserted between anti-ferromagnetically aligned spins, thus reducing the long-ranged antiferromagnetic order to a quasi-long range order \cite{main}. Removing the holes and doublons, which can be accomplished by the string order $\langle \mathfrak{S}_{ij} \rangle=\langle S^z_i(\prod_{i<k<j} P_k)S^z_j \rangle$ with on-site fermion parity operator $P_k$, recaptures anti-ferromagnetic correlations \cite {main,main39,main54}. Since the operators at the ends of the string, $S^z_i$ and $S^z_j$, flip under $R_x$ and the bulk has well-defined fermion parity $P$, a non-vanishing string order that extends across the entire chain with open boundaries implies localized edge modes with $\langle S^z_1\rangle, \langle S^z_L\rangle\neq0$. Such edge modes emerge only if the spins are quasi-long ranged ordered, indicating that the non-trivial topology is contingent on the bulk being gapless.

 \par
\begin{figure}
		\includegraphics[width=\linewidth]{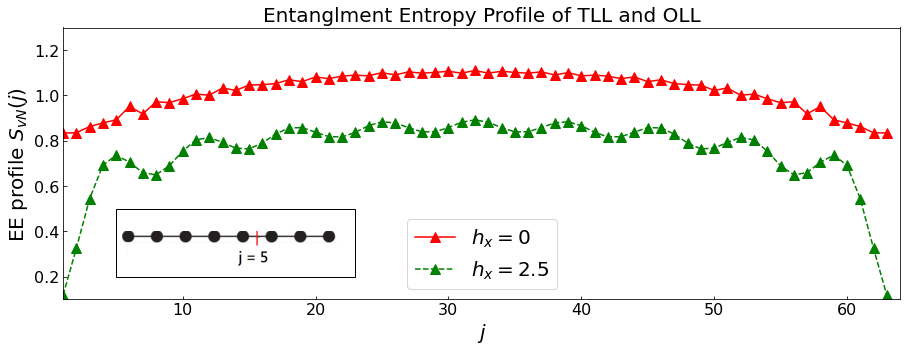}
		\caption{The EE profile of the TLL ($h_x=0$) and the OLL ($h_x=2.5$) in the Ising-Hubbard chain. As a function of the left subsystem size $j$, indicated inset, the former shows no oscillations whereas the latter shows clear oscillations that can be checked to correspond to twice the Fermi wavevector.
}
\label{fig:TLL_EE}
\end{figure}

\begin{figure}
		\includegraphics[width=\linewidth]{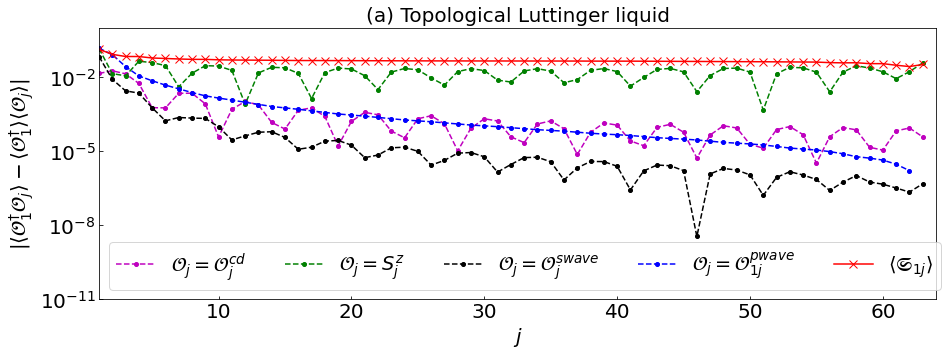}
		\includegraphics[width=\linewidth]{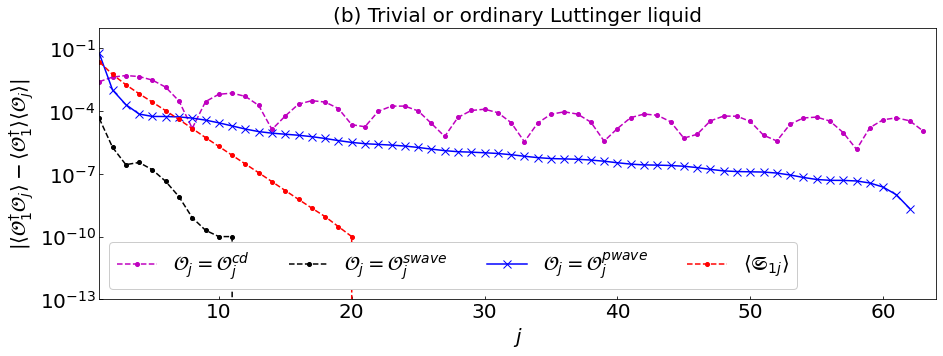}

\caption{Correlations in the TLL (a) and the OLL (b). In the TLL, most correlations of local operators are quasi-long-ranged and oscillating, p-wave correlations are quasi-long-ranged and lack oscillations and the string order is long-ranged and non-oscillating. In the OLL, string order decays exponentially, charge correlations are quasi-long-ranged with Friedel oscillations while p-wave correlations are quasi-long-ranged and non-oscillating. In both figures, the only local correlation with non-vanishing disconnected part $\langle \mathcal{O}_i^\dag \rangle \langle \mathcal{O}_j  \rangle$ is $\mathcal{O}_{j}=\mathcal{O}_{j}^{c d}$ due to conserved $R_x$ and total particle number. \label{fig:correlations}}
\end{figure}

 We now fix the parameters $t_c=0$, $U=5$, $\mu=0.5,$ and $t=J_z=1.0$ and tune $h_x$ to generate Figs.~\ref{fig:TLL_EE} and \ref{fig:correlations}. At $h_x=0.0$, the model yields a TLL ground state while a larger value, $h_x=2.5$, results in an OLL.
  
Fig.~\ref{fig:TLL_EE} shows the EE of a finite chain $S_{vN}(j)$ as a function of left subsystem size $j$ for the TLL and the OLL. The EE profile of OLL shows clear Friedel oscillation induced by the boundary. However, such oscillations are absent in the TLL. 

To examine whether the absence of EE oscillations is related to any correlations, we plot the absolute value of various correlation functions in the TLL and the OLL in Fig.~\ref{fig:correlations}.
In the TLL , all correlators of local operators show quasi-long ranged order whereas the string order has nearly constant magnitude and hence, dominates at large distances. Moreover, all the local correlators except correlations between p-wave Cooper pairs show clear oscillations. The absence of oscillations in the latter -- which was also pointed out in Ref.~\cite{main} -- suggests that EE profile lacks oscillations because the lowest energy excitations are p-wave Cooper pairs. However, an inspection of correlations in the OLL, which contains EE oscillations, indicates otherwise. Indeed, Fig.~\ref{fig:correlations}(b) shows that p-wave correlations still lack oscillations even though the EE oscillates in Fig.~\ref{fig:TLL_EE}.

\begin{figure}
		\includegraphics[width=\linewidth]{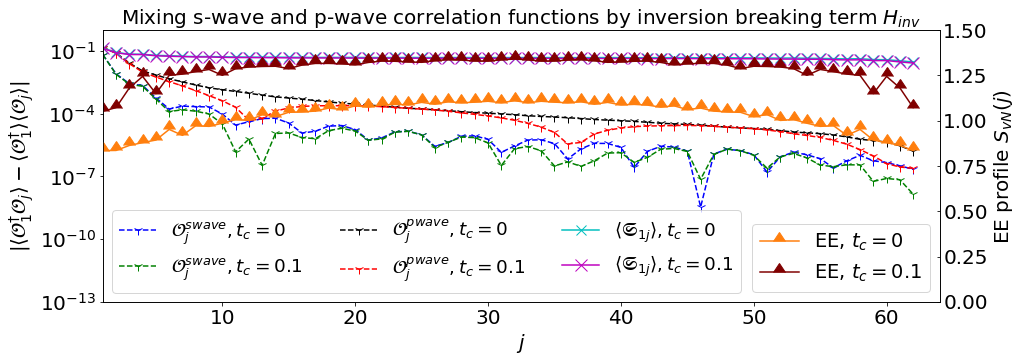}
		\caption{Correlations, EE and string order with ($t_c=0.1$) and without ($t_c=0$) inversion symmetry breaking in $H_1$. With broken inversion, s- and p-wave Cooper pairs mix and both correlations show oscillations whereas p-wave correlations are non-oscillating when inversion is preserved. In both cases, the EE profile and the string order lack oscillations.}
		\label{fig:inverbreak-TLL} 
\end{figure}

To unambiguously disentangle the absence of oscillations in p-wave correlations from the absence in EE, we break inversion symmetry weakly in the TLL by setting $t_c=0.1$ in $H_\textrm{inv}$. This term is expected to mix s-wave and p-wave Cooper pairs but leave the TLL robust since it is protected by $R_x$ symmetry. Indeed, as shown in Fig.~\ref{fig:inverbreak-TLL}, p-wave pairing correlations also show oscillations when $t_c\neq0$ while the string order and the EE profile remain oscillation-free. Therefore, the absence of EE oscillations in the TLL is a more fundamental property than the existence of local operators with non-oscillating correlations.

 \par We next investigate the connection between the absence of oscillations in the EE profile and the presence of protected edge modes. To do so, we consider a gapless phase that also has edge modes but is \emph{not} a TLL, namely, a doped spin-1 Haldane chain \cite{DHLeespt}. We briefly describe the model below before presenting our results on it.


\emph{Doped Haldane}.-- To facilitate a comparison with the Ising-Hubbard chain, we consider a fermionic realization of the doped spin-1 Haldane chain as a t-J type ladder Hamiltonian, $H_2=H_t+H_J$: 

\begin{eqnarray} \label{eq:haldane} 
	H_{t}&=&-t\sum_j\sum_{\alpha=1,2}\sum_{s}(c^\dag_{\alpha,j+1,s}c_{\alpha,j,s}+h.c.)\\ 
	H_J&=&J\sum_{i}\sum_{\alpha=1,2}\vec{S}_{\alpha,i}\cdot\vec{S}_{\alpha,j+1}+J_{\bot}\sum_{j}\vec{S}_{1,j}\cdot \vec{S}_{2,j+1}
\end{eqnarray}
with the constraint that there is at most one electron per site \cite{DHLeespt}. Here $\alpha=1,2$ labels the two chains. In the regime $J_\perp<0$, $J>0$, $|J_\perp|\gg J,|t|$, electrons on sites connected by a rung effectively form a spin-1 boson that can hop to neighboring rungs with amplitude $O(t^2/|J_\perp|)$ and is coupled anti-ferromagnetically to its neighbors. If $t=0$, the ground state is the well-known Haldane SPT with a bulk gap and spin-1/2 excitations on the edge. In the opposite limit, $t^2/|J_\perp|\gg J$, Ref.~\cite{DHLeespt} showed that the ground state can be understood as a spin-1 anti-ferromagnet but on a ``squeezed" lattice, i.e., a lattice with holes removed, and has gapless charge excitations. Remarkably, the edge modes survive this limit, which is reminiscent of the TLL derived from the Ising-Hubbard model. However, the physics of the two phases is fundamentally different: the edge modes from the gapped Haldane chain survive gaplessness in the doped Haldane chain whereas gaplessness is \emph{necessary} for edge modes to appear in the TLL.

\begin{figure}	
		\includegraphics[width=\linewidth]{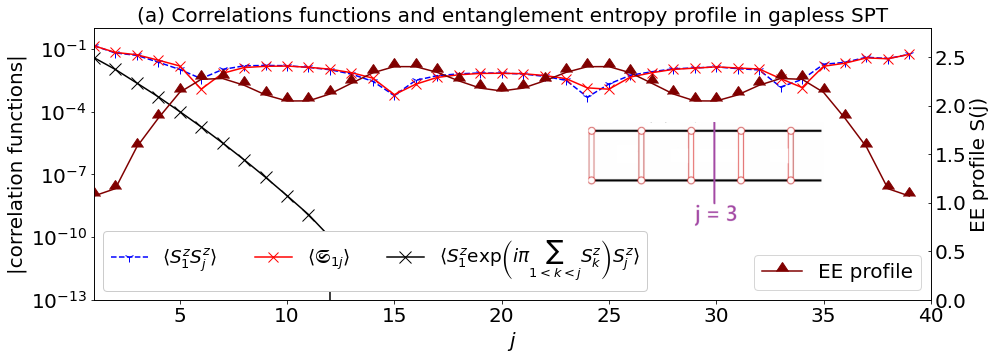}
		\begin{flushleft}
		\includegraphics[width=0.95\linewidth]{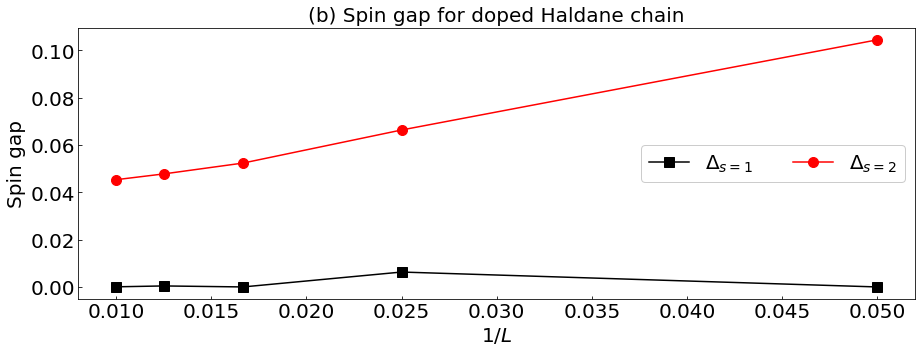}
		\end{flushleft}
		\caption{(a) Correlations, string orders and EE profile in the doped Haldane chain (depicted inset along with the entanglement cut). The EE profile, $S^z$-correlations and string order $\langle\mathfrak S_{1j}\rangle$ show oscillations, while the string order from the Haldane SPT phase, $\left\langle S_1^z\exp\left(i\pi\sum_{1<k<j}S_k^z\right)S_j^z\right\rangle$ decays exponentially. (b) Finite-size scaling of spin gaps. The gap for $\Delta S^z=1$ ($\Delta S^z=2$) vanishes (is finite) in the thermodynamic limit, which is consistent with a spin-1/2 zero mode at each end but not with spin-1 excitations being gapless.
		\label{fig:doped Haldane}}

\end{figure}

In Fig.~\ref{fig:doped Haldane}, we show results for $t=1$, $J=0.3$, $J_\bot=-3$ and $10\%$ doping, some of which were presented in Ref.~\cite{DHLeespt} for $5\%$ doping which is computationally more taxing. In Fig.~\ref{fig:doped Haldane}(a), we plot the correlations along chain $\alpha=1$ as well as the EE as a function of the entanglement cut position as indicated, while  Fig.~\ref{fig:doped Haldane}(b) provides evidence that edge modes are indeed present. Interestingly, the EE profile shows clear oscillations, proving that the mere presence of edge modes cannot suppress EE oscillations in a gapless phase. Fig.~\ref{fig:doped Haldane} also shows that the string order along one of chains, $\langle \mathfrak{S}_{ij}\rangle=\langle S^z_{\alpha=1,i}(\prod_{i<k<j} P_{\alpha=1,k})S^z_{\alpha=1,j} \rangle$ and has Friedel oscillations with a similar wavelength as the EE profile. These observations lend further support to the hypothesis that the absence of EE oscillations is tied to the presence of a long-ranged string order.

\par $Discussion$---
Having provided evidence that the absence of EE oscillations is closely associated with the string order, we now provide an intuitive picture for this connection.

\begin{figure}
		\includegraphics[width=\linewidth]{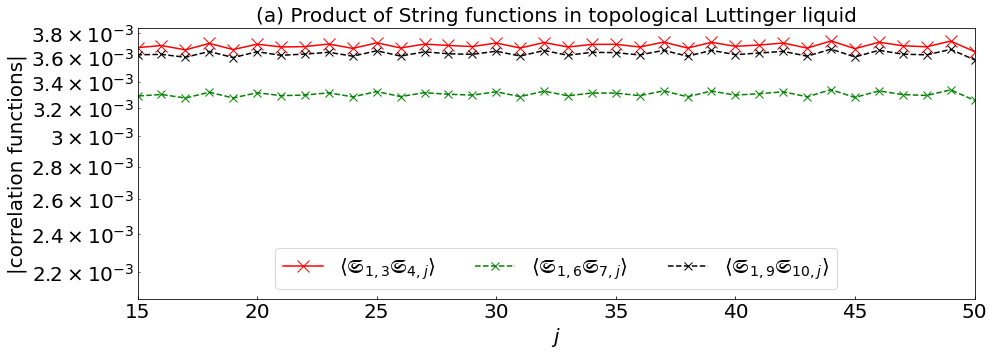}
		\includegraphics[width=\linewidth]{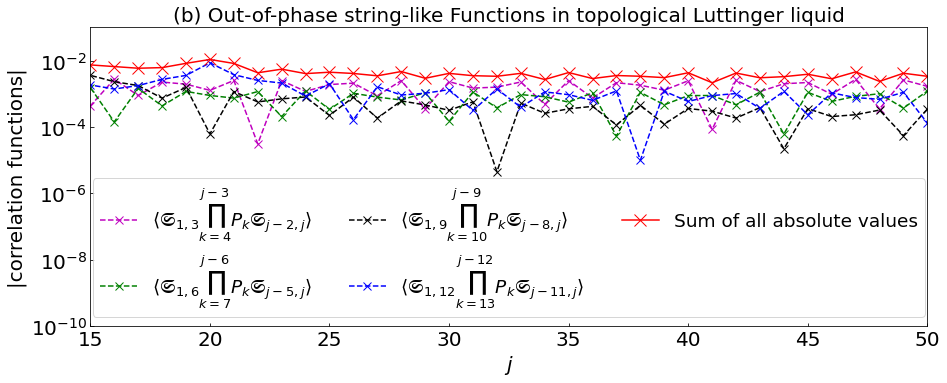}

\caption{(a) Absolute value of several products of two strings showing long-range order with negligible fluctuations inherited from the single string order $\mathfrak S_{1j}$ in Fig. \ref{fig:TLL_EE}. (b) String products separated by fermion parity insertions show larger fluctuations. However, the fluctuations are out-of-phase, so the sum of their absolute values is significantly smoother.}
		\label{fig:string-like}
\end{figure}

The EE of a region can be understood as the averaged behavior of all correlation functions in that subsystem. For example, the second Renyi entropy is precisely the (negative logarithm of the) mean square expectation value of all operators in the region, and the Von Neumann entropy can be intuitively understood similarly. To see how this idea results in the non-oscillating EE profile, we show the behavior of more long-ranged string-like functions with non-vanishing magnitude in TLL. In particular, we show the 
products of two adjacent strings, $\langle \mathfrak{S}_{1,i}\mathfrak{S}_{i+1,j}\rangle$, in Fig.~\ref{fig:string-like}(a) and pairs of short strings separated by a string of fermion parity, $\left\langle \mathfrak{S}_{1,i} \prod^{j-i}_{k=i+1} P_k\mathfrak{S}_{j-i+1,j}\right\rangle$  in Fig.~\ref{fig:string-like}(b). The latter can also be written as a single string flanked by local operators on either side, $\left\langle \left(S^z_1\prod_{k=2}^{i-1}P_k\right)\mathfrak S_{i,j-i+1}\left(\prod_{k=j-i+2}^{j-1}P_{k}S_j^z\right)\right\rangle$. All these correlations inherit long-range order from the fundamental string order $\mathfrak S_{i,j}$. As the subsystem size $j$ varies, the products of adjacent strings do not fluctuate while the products of separated short strings fluctuate with unrelated phase. Thus, in the presence of long-range string order -- a defining property of the TLL -- the EE receives dominant contributions from the non-oscillating string order and the large number of two-, three- and higher string-products which either lack spatial variations or exhibit fluctuations that cancel out upon summation. As a result, the EE profile lacks oscillations too.

Finally, we study the EE profile of two other TLLs, a charge-conserving topological superconductor \cite{MEMTLL} and fermions with attractive triplet pairing interactions \cite{BergTLL}. The respective Hamiltonians are:
\begin{eqnarray}
H_{\textrm{TSC}}&=&\sum_{j,s} -A_{j}^{s}+g(A_{j}^{s})^{2}+\sum_j B_{j}+g B_{j}^{2} \label{eq:MEMs}\\
H_{\textrm{triplet}}&=&-t\sum_{j,s}A^s+U\sum_j\Delta_{j}^\dag  \Delta_{j}\label{eq:triplet}
\end{eqnarray}
where $B_{j}=c_{j,\uparrow}^{\dagger} c_{j+1,\uparrow}^{\dagger} c_{j,\downarrow} c_{j+1,\downarrow}+\textrm{h.c.}$ and $\Delta_{j}=c_{j, \uparrow} c_{j+1, \downarrow}+\textrm{h.c.}$. The ground state of $H_\textrm{TSC}$ is a TLL at $g=0.9$ at half-filling \cite{MEMTLL}, while the ground state of $H_\textrm{triplet}$ is a TLL at $t=-U=1.0$ at density $n=\frac{N_\uparrow+N_\downarrow}{L}=0.2$ \cite{BergTLL}. $H_\textrm{triplet}$ is also known to have long-range string order without Friedel oscillations \cite{main, enriched}. 

\begin{figure}	
		\includegraphics[width=\linewidth]{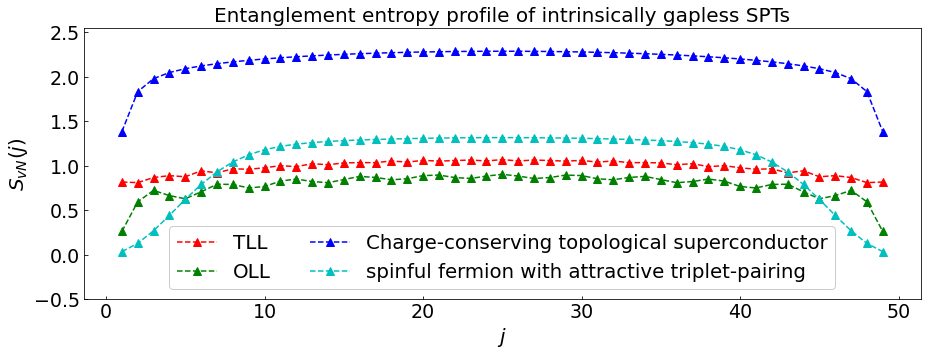}
		\caption{ EE profiles of two other TLLs, the charge conserving topological superconductor and fermions with spin-orbit coupling and triplet pairing defined in Refs.~\cite{MEMTLL} and \cite{BergTLL}, respectively, alongside the Ising-Hubbard TLL and OLL from Fig.~\ref{fig:TLL_EE}. EE oscillations are clearly absent in all TLLs.
		  }
		\label{fig:EEshort}
    
\end{figure}

For both models, we perform DMRG calculations with $222$ sweeps and $\chi\approx500$, resulting in truncation error $\epsilon\approx10^{-10}$ for system size $L=50$. Fig.~\ref{fig:EEshort} shows the results. Clearly, the absence of Friedel oscillations in the EE profile is prevailing in these models too, further supporting our claim that TLLs generically lack Friedel oscillations in the EE. 


\par $Conclusion$-- We have shown that the EE profile of TLLs lacks Friedel oscillations and argued that this behavior is closely related to the long-ranged string order present in this phase. Thus, it forms an unbiased signature of the phase as it does not require precise knowledge of the string order parameter. First, we noted that p-wave pairing correlations were the only local correlations that were non-oscillating in the Ising-Hubbard TLL. We then broke inversion symmetry in the model, which induced oscillations in the p-wave correlations due to mixing with s-wave pairing while the EE profile and string order remained non-oscillating. This ruled out non-oscillating p-wave correlations -- and all non-oscillating local correlations by extension -- as the cause of the absence of EE oscillations in the Ising-Hubbard TLL, while indicating a deeper connection to the string order. Next, we investigated the role of edge modes in suppressing EE oscillations by studying a gapless SPT, the doped spin-1 Haldane chain, which also has edge modes. However, EE oscillations exist in this phase, implying that the mere presence of edge modes cannot suppress EE oscillations. Finally, we described an intuitive picture for the absence of EE oscillations in TLLs in terms of expectation values of composite strings. Specifically, we argued that the EE in TLLs is dominated by the expectation values of products of strings, all of which inherit long-range order from the fundamental string. Since long-ranged string order is a defining property of TLLs, the absence of Friedel oscillations in its EE profile likely is too. It would be interesting to ask whether EE oscillations are also absent in gapless topological phases protected by nonlocal symmetries, e.g., the triplet paired phase in the mass imbalanced Hubbard chain that has a string order protected by global inversion but no edge modes \cite{HegaplessSPT}. We leave this question for future work.


\acknowledgements
We acknowledge financial support from the National Science Foundation grant DMR 2047193.

\bibliographystyle{apsrev4-2}
\bibliography{prbref}

\end{document}